# Resilience assessment framework for cyber-physical distribution power system based on coordinated cyber-physical attacks under dynamic game

Yulin Liu, Zhaojun Ruan, and Libao Shi, *Senior Member, IEEE*

*Abstract*—Owing to the advanced communication networks and intelligent electronic devices, the cyber-physical distribution systems (CPDSs) possess the capability to perform flexible economic dispatch and achieve rapid self-healing from extreme events. Meanwhile, the deep integration of cyber and physical systems makes CPDS vulnerable to coordinated cyber-physical attacks. In this paper, a resilience assessment framework for the CPDS under coordinated cyber-physical attacks is proposed to investigate the impact of the coordinated attacks on load loss and service restoration in CPDS. First, a three-stage defender-attacker-defender dynamic game model considering fake base station (FBS) and physical attacks for CPDS is established, aiming at seeking the optimal defense resource deployment strategy to enhance the resilience of the CPDS. The physical attack is launched to cause faults on the power lines, and the FBS attack is employed to interrupt the service of wireless cellular network to hinder the self-healing process of the CPDS. The lognormal shadowing model and search theory are applied to quantitatively describe the process of the coordinated cyber-physical attacks. Further, the constructed three-stage dynamic game model is equivalently recast as a tri-level max-min-max optimization model, which is solved using column-and-constraint generation combined with enumeration method. Finally, the effectiveness of the proposed resilience assessment framework and solution strategy is demonstrated by conducting simulation analysis on the modified IEEE 33-node CPDS and a real-world 47-node CPDS in China.

*Index Terms*—Resilience assessment, cyber-physical distribution power system, fake base station, coordinated cyber-physical attack, dynamic game.

## I. INTRODUCTION

With the advancement of communication technology and the widespread application of intelligent terminal devices (IEDs), the traditional distribution networks have gradually evolved into cyber-physical distribution systems (CPDSs) [1]. Through IEDs such as smart meters, remote terminal units (RTUs), and remote-controlled switches (RCSs), the feeder automation (FA) system can achieve fault location, isolation, and service restoration [2], thereby enhancing the resilience of the CPDS. However, at the same time, owing to the lack of protection in the cyber system and the stealth features of cyber attacks, the cyber side of the CPDS has become the primary target of attackers [3]. In particular, some cyber attacks in CPDS, such as false data injection (FDI) attacks, denial-of-service (DoS) attacks, and replay attacks, can perform in conjunction with physical attacks [4], further amplifying the effects of damaging systems, including interrupting key power supply services and slowing down the system's ability to recover from failures, that is, reducing the resilience of the CPDS. Therefore, it is of great practical significance to perform resilience assessment for the CPDS under the coordinated cyber-physical attacks.

So far, many studies have been conducted on the cyber-physical security of CPDS under various malicious attacks. It should be noted that some studies focused on the risk assessment of CPDS, aiming at identifying the vulnerable cyber and physical nodes as well as demonstrating the risk level under different attacks. A risk assessment model for FA system in active distribution network under cyber attacks was proposed in [5]. In the model, the search theory was employed to calculate the probability of the attacker targeting a specific RTU, and a Bayesian attack graph was constructed to model the probability of the attacker successfully exploiting cyber vulnerabilities. Similarly, a risk assessment for CPDS was conducted based on risk propagation model and analytic hierarchy process method [6]. Further, a two-layer attack-defense model [7] was constructed to assess the risk of load redistribution attacks (LRAs) on conservation voltage reduction (CVR) in a three-phase unbalanced active distribution network. The corresponding results demonstrated that the LRAs on CVR could lead to voltage limit violations at certain nodes. In addition, there were also studies that established optimization model between defenders and attackers based on game theory, and proposed optimal mitigation strategy. For instance, a tri-level defender-attacker-defender (DAD) model [8] was proposed to address cyber attacks on Volt/VAR optimization in distribution management system. The DAD model was solved with the nested column-and-constraint generation (C&CG) algorithm and identified the most essential smart meters that should be hardened. Further, a robust strategy [9] was introduced to mitigate load altering attacks in CPDS by leveraging soft open points, and a two-stage distributed robust chance-constrained optimization model was constructed to ensure the robustness of the strategy against the ambiguity of the load empirical probability distributions.

Y. Liu is with the Department of Electrical Engineering, Tsinghua University, Beijing, 100084, P. R. China and the Tsinghua Shenzhen International Graduate School, Tsinghua University, Shenzhen 518055, P. R. China (e-mail: liu-yl22@mails.tsinghua.edu.cn).
Z. Ruan and L. Shi are with the Tsinghua Shenzhen International Graduate School, Tsinghua University, Shenzhen 518055, P. R. China (e-mail: rzj23@mails.tsinghua.edu.cn; shilb@sz.tsinghua.edu.cn).

In recent years, there has been rapid development in wireless communication network technologies and their applications in power system. In particular, 5G cellular networks have been widely studied and applied in the CPDSs, providing significant high-bandwidth and low-latency services [10]. Several studies have investigated the deployment architectures of 5G networks to address the performance requirements of power system communications. In [11], an efficient radio resource management method for 5G radio access networks (RANs) was proposed, leveraging deep reinforcement learning to meet the performance requirements of smart grid self-healing applications. In [12], a 5G RAN slicing strategy was introduced to facilitate the participation of virtual power plants in frequency regulation services. However, wireless communication networks are inherently more vulnerable to cyber attacks than wired networks due to their broadcast nature [13], prompting further research into malicious attacks on such systems. For instance, an attack strategy targeting wireless smart grid networks [14] was proposed, utilizing spoofing and jamming techniques with optimal power distribution to maximize the adversarial effects. In addition, a tri-level DAD model [15] was developed to address coordinated attacks on base stations (BSs) and power lines, incorporating a base station energy adjustment strategy to enhance the resilience of CPDS.

Fake base stations (FBSs) represent a critical security threat to wireless networks, as these illegal programmable radio devices impersonate legitimate base stations to facilitate network eavesdropping, FDI attacks, and DoS attacks [16]. Given the severity of these threats, both industry and academia have prioritized the development of robust security measures to counteract FBS attacks. The technical report 3GPP TR 33.809 [17] explored cryptographic solutions, including digital signature-based, certificate-based, and identity-based approaches, as potential countermeasures. In [18], a theoretical model of the received signal strength distribution for user equipment was developed, leading to the proposal of a FBS signal detection method based on received signal strength thresholds. In addition, a blockchain-based system information protection framework [19] was developed to address the potential attacks and failures, which was theoretically analyzed and demonstrated to offer superior defense capabilities against FBS attacks.

The existing studies on FBS primarily concentrated on signal detection methods and system information protection, thereby enhancing the security level of wireless terminal devices. However, as the deployment of wireless communication and control devices in distribution networks continues to expand, there remains a notable gap in the literature regarding the resilience assessment of CPDS under the threat of coordinated attacks that combine FBS with physical attacks. Addressing this gap is critical to ensuring the robustness and reliability of CPDS in the face of increasingly sophisticated and multifaceted security threats. In this paper, a resilience assessment framework for the CPDS under coordinated cyber-physical attacks is proposed to investigate the impact of the coordinated attacks on the resilience of CPDS. The main contributions of this paper are summarized in following three-fold:

(1) A three-stage DAD dynamic game model considering coordinated cyber-physical attacks for CPDS is proposed to assess the system resilience. The dynamic game model is equivalently recast as a tri-level optimization model, which is solved by C&CG algorithm and enumeration method.

(2) The FBS attack is first considered in the attack model for CPDS, and the principle and implementation process of FBS are analyzed elaborately. The lognormal shadowing model is employed to calculate the signal strength received by wireless RCSs from the BSs, and an approximate calculation method based on piecewise polynomial fitting is proposed to calculate the attack probability, which is a transcendental integral function, and its upper error bound is proved as well.

(3) The multi-stage service restoration process with cyber-physical coupling is introduced into the service restoration model, taking into account the impact of attacked RCSs on the physical-side recovery process.

## II. PROBLEM FORMULATION

This paper aims to investigate a category of coordinated cyber-physical attacks targeting CPDS integrated with wireless cellular networks, in which attackers launch physical attacks to induce faults in the power lines while simultaneously deploying FBS attacks on the cyber side to sever the main station's control over RCSs. To cope with this multifaceted threat, it is necessary to elaborately explore the resilience enhancement strategies against such coordinated attacks. In view of this, a three-stage dynamic game DAD framework is established, as shown in Fig. 1.

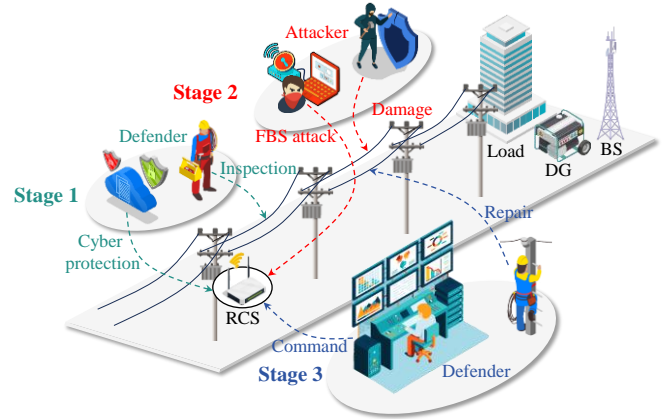

**Fig. 1.** Schematic diagram of three-stage dynamic game DAD framework.

In Stage 1, the defender allocates physical resources to enhance the inspection intensity of selected power lines and deploys cyber resources to strengthen the resilience of specific RCSs against FBS attacks. In Stage 2, the attackers launch physical attacks to induce faults in power lines while simultaneously identifying optimal locations to deploy FBS for initiating cyber attacks. The resulting loss of control over RCSs due to FBS attacks disrupts the self-healing process of the CPDS, thereby exacerbating the overall damage. In Stage 3, the system operator implements multi-stage recovery operations to restore the CPDS, including rapid fault isolation, service restoration through RCSs, on-site control of disconnected switches, and repair of faulty power lines. The players in this three-stage dynamic game are the defender, representing CPDS decision-makers focused on enhancing system resilience in Stage 1 and Stage 3, and the attacker, representing a malicious sabotage team in Stage 2. The fundamental assumptions of the three-stage dynamic game model are outlined as follows:

(1) The attacker team possesses comprehensive knowledge of both the power system and wireless communication infrastructure, enabling them to access the topology, operational data of the distribution network, and the locations of BSs.

(2) Given that the probability of faults in power lines and disruptions in wireless cellular network under normal operating conditions is significantly lower than in attacked scenarios, the probability of such faults and RCS disconnections under normal circumstance is assumed to be zero.

(3) The duration of each stage in the recovery process of the CPDS is predetermined and remains constant.

*A. The wireless cellular network and FBS attack*

In wireless cellular networks, the BS serves as the primary gateway for wireless IEDs to access the communication network, acting as the bridge between the IEDs and the core network. Within 5G cellular networks, the IED establishes a connection to the BS through the radio resource control (RRC) layer and to the core network via non-access stratum (NAS) layer [20], as shown in Fig.2. The RRC layer connection process initiates with the BS periodically broadcasting system information block (SIB) messages on a specific downlink frequency. These messages contain critical cell selection criteria and network identities, including the PLMN ID and the cell ID, where the PLMN ID uniquely identifies the network provider and the cell ID identifies the BS broadcasting the SIB message. Subsequently, the IED and BS complete the RRC connection through a three-way handshake: (i) the IED sends the RRC setup request to the BS; (ii) upon receiving the request, the BS accepts the connection and sends the RRC setup message to the IED; and (iii) the IED responds with the RRC setup complete message to finalize the connection. The NAS layer connection process follows a similar three-way handshake mechanism, involving the NAS registration request, the NAS authentication request, and the NAS authentication response. Notably, when multiple BSs are within range, the IED is configured to connect to the BS with the strongest received SIB signal strength, a fundamental principle of the cellular network's cell selection mechanism [21].

In the described framework, attackers with expertise in wireless communication can intercept the SIB information from a legitimate BS and subsequently launch a FBS attack using a wireless transceiver and a computing unit, such as a laptop, an Intel NUC or a Raspberry Pi. Leveraging the characteristic of IED to connect to the BS with the strongest received SIB signal strength, the FBS is designed to broadcast forged SIB messages with elevated signal strength or from a position closer to the IEDs, thereby deceiving the IEDs into initiating RRC setup requests. Once the RRC layer connection is established using forged information, the FBS intercepts the NAS registration request from the IEDs and deliberately withholds it from the core network, halting the subsequent communication process [22]. As shown in Fig.2, the FBS neither releases the compromised IEDs nor transmits any further messages, resulting in the interruption of communication services for the attacked IEDs.

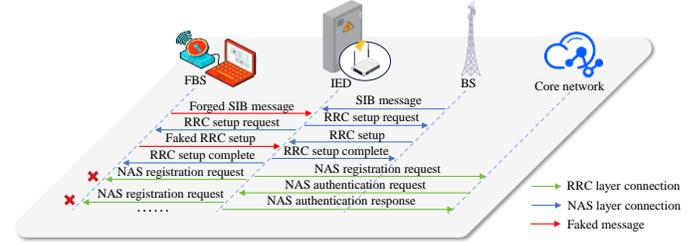

**Fig. 2.** The process of wireless IED connecting to the BS and the FBS attack mechanism.

*B. Modelling of three-stage defense and attack scenarios*

1) Stage 1: modeling defense resource deployment

In Stage 1, the defenders proactively allocate defense resources to both the cyber and physical sides of the CPDS prior to the event with the objective of enhancing the system's resilience against coordinated attacks. On the physical side, these resources involve deploying power grid defenders to intensify the inspection and monitoring of power lines. The probabilities of detecting and deterring physical attacks on the power lines can be quantitatively modelled using search theory [23], as expressed by the following formulations:

$$p_{ij,a} = e^{-l_{ij}/\zeta_a}, ij \in O \quad (1a)$$
$$p_{ij,b} = e^{-l_{ij}/\zeta_b}, ij \in O \quad (1b)$$

where $l_{ij}$ represents the length of power line $ij$, and $O$ represents the set of power lines. $\zeta_a$ denotes the normal inspection intensity, while $\zeta_b$ corresponds to the enhanced inspection intensity. In addition, $p_{ij,a}$ and $p_{ij,b}$ denote the probabilities of detecting and deterring physical attacks on the power line $ij$ under the normal and enhanced inspection intensities, respectively.

Consequently, the probability of the successful physical attack on the power line $ij$ can be expressed as follows:

$$p_{ij} = A_{p,ij}\left(1 - \left(p_{ij,a} - D_{p,ij}(p_{ij,a} - p_{ij,b})\right)\right), ij \in O \quad (2)$$

where $p_{ij}$ denotes the probability of the successful physical attack on the power line $ij$, $A_{p,ij}$ is a binary variable indicating whether the power line $ij$ is subjected to a physical attack, while $D_{p,ij}$ is a binary variable indicating whether the power line $ij$ is reinforced with enhanced inspection measures.

On the cyber side, defender can protect the RCSs against the FBS attacks by deploying the detection program [18], [22], [24] or strengthening the verification mechanisms for BSs [19], [25]. These protective measures ultimately reduce the success probability of FBS attacks, which can be expressed as follows:

$$p_w = p_{a,w}(1 - D_{c,w}p_{d,w}), w \in W \quad (3)$$

where $W$ denotes the set of RCSs, $p_w$ denotes the overall success probability of a FBS attack on the RCS $w$, and $p_{a,w}$ represents the probability that the FBS sends the SIB message to the RCS $w$ and achieves the strongest signal strength. In addition, $D_{c,w}$ is a binary variable indicating whether the RCS $w$ is protected, and $p_{d,w}$ denotes the probability that the protection mechanism effectively mitigates the FBS attacks.

In practical applications, the defense resources allocated to both the physical and cyber sides are constrained, as expressed by the following limitations:

$$\sum_{ij \in O} D_{p,ij} \leq N_{d,p} \quad (4a)$$
$$\sum_{w \in W} D_{c,w} \leq N_{d,c} \quad (4b)$$

where $N_{d,p}$ denotes the maximum number of power lines that can be protected, and $N_{d,c}$ denotes the maximum number of RCSs that can be protected against FBS attacks.

2) Stage 2: modeling coordinated cyber-physical attack

The attackers launch physical attacks on specific power lines and simultaneously initiate FBS attack to disrupt the self-healing process of the CPDS by rendering some RCSs uncontrolled. In real-world scenarios, the number of lines that attackers can target is limited, as expressed in (5):

$$\sum_{ij \in O} A_{p,ij} \leq N_{a,p} \quad (5)$$

where $N_{a,p}$ represents the maximum number of power lines that can be attacked.

The location of the FBS has a significant impact on the choice of the RCSs to attack and the success probability of the FBS attack. Therefore, experienced attackers optimize the deployment location of the FBS within a specific geographical range. To simplify the model, the continuous geographical area of the CPDS is discretized into grid positions, as shown in Fig.3, and the location of the FBS will be selected from the grid intersections with coordinates $(X_{FBS}, Y_{FBS})$.

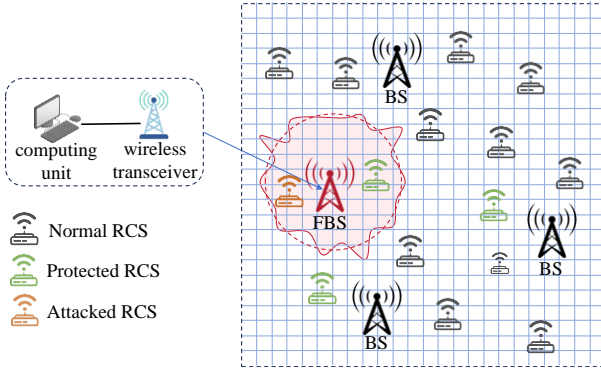

**Fig. 3.** Schematic diagram of the FBS location choice.

Assuming there are $N + 1$ BSs near the RCS $w$, with the first $N$ being legitimate BSs and the $(N + 1)_{th}$ being the FBS deployed by the attackers, the received SIB signal strengths from these BSs are $\{S_1, S_2, ..., S_{N+1}\}$. According to the lognormal shadowing model [26], the signal strength $S_k$ received by the RCS $w$ is a random variable related to the propagation distance and follows a normal distribution:

$$S_k = S(d_0) - 10\eta \lg\left(\frac{d_{k,w}}{d_0}\right) + X_k \quad (6)$$

where $S(d_0)$ is the reference signal strength at the distance of $d_0$ from the BS, $\eta$ is the path loss coefficient, $d_{k,w}$ is the distance between the RCS $w$ and the BS $k$, and $X_k$ is the random attenuation of path loss due to the shadowing effect. $X_k$ follows a normal distribution $X_k \sim \mathcal{N}(0, \sigma_k^2)$, where $\sigma_k$ is the standard deviation. Thus, the random variable $S_k$ also follows a normal distribution:

$$S_k \sim \mathcal{N}\left(S(d_0) - 10\eta \lg\left(\frac{d_{k,w}}{d_0}\right), \sigma_k^2\right) \quad (7)$$

When multiple BSs are nearby, the RCS will connect to the BS with the strongest received SIB signal strength. Therefore, the probability $p_{a,w}$ of the FBS (the $(N + 1)_{th}$ BS) sending the SIB message to the RCS $w$ and achieving the strongest signal strength is computed as:

$$p_{a,w} = p\left\{S_{N+1} > \max_{1 \leq k \leq N} S_k\right\} = \int_{-\infty}^{+\infty} f_{N+1}(s) \prod_{k=1}^{N} F_k(s) ds \quad (8)$$

where $f_{N+1}(s)$ is the probability density function (PDF) of the received signal strength from the FBS, and $F_k(s)$ is the cumulative distribution function (CDF) of the received signal strength from the legitimate BS $k$. The physical meaning of (8) is that the probability of the received signal strength being $s$ is $f_{N+1}(s)ds$, and in this case, the probability of the FBS having the strongest signal strength is $\prod_{k=1}^{N} F_k(s)$. Since the variable $s$ can take any value, the overall probability is the integral of $f_{N+1}(s) \prod_{k=1}^{N} F_k(s)$.

Given that the signal strengths follow normal distributions $S_k \sim \mathcal{N}(\mu_k, \sigma_k^2)$ and $S_{N+1} \sim \mathcal{N}(\mu_{N+1}, \sigma_{N+1}^2)$, let $\tau \triangleq \frac{s - \mu_{N+1}}{\sigma_{N+1}}$ and transform $f_{N+1}(s)$ into the standard normal distribution form:

$$p_{a,w} = \int_{-\infty}^{+\infty} f_{N+1}(\sigma_{N+1}\tau + \mu_{N+1}) \cdot$$
$$\prod_{k=1}^{N} F_k(\sigma_{N+1}\tau + \mu_{N+1}) d(\sigma_{N+1}\tau + \mu_{N+1})$$
$$= \int_{-\infty}^{+\infty} \frac{1}{\sqrt{2\pi}\sigma_{N+1}} e^{-\frac{(\sigma_{N+1}\tau + \mu_{N+1} - \mu_{N+1})^2}{2\sigma_{N+1}^2}} \cdot \quad (9)$$
$$\prod_{k=1}^{N} F_k(\sigma_{N+1}\tau + \mu_{N+1}) \sigma_{N+1} d\tau$$
$$= \frac{1}{\sqrt{2\pi}} \int_{-\infty}^{+\infty} e^{-\frac{\tau^2}{2}} \prod_{k=1}^{N} \Phi\left(\frac{\sigma_{N+1}}{\sigma_k}\tau + \frac{\mu_{N+1} - \mu_N}{\sigma_k}\right) d\tau$$

where $\mu_k = S(d_0) - 10\eta \lg\left(\frac{d_{k,w}}{d_0}\right)$, $\mu_{N+1} = S(d_0) - 10\eta \lg\left(\frac{d_{N+1,w}}{d_0}\right)$, and $\Phi(\cdot)$ denotes the CDF of the standard normal distribution.

According to the 3-sigma principle of the normal distribution, if the received signal strengths from the FBS and other legitimate BSs satisfy $\left|\mu_{N+1} - \max_{1 \leq k \leq N}(\mu_k)\right| \geq 3\sigma_{N+1}$, the relationship between the signal strengths can be considered deterministic, effectively ignoring the impact of the shadowing effect.

3) Stage 3: modeling service restoration process after the coordinated attacks

After the fault occurred at $T_f$ owing to the coordinated cyber-physical attacks, the multi-stage service restoration process of the CPDS mainly involves the following key points: (i) at $T_{r,1}$, the defenders utilize the RCSs to conduct the fault isolation and load transfer operation; (ii) some RCSs may be out of service in the first stage of service restoration owing to the FBS attack. Therefore, at $T_{r,2}$, the power grid defender will control the disconnected switches on-site to complete the fault isolation and load transfer operation; and (iii) at $T_{r,3}$, the faults are repaired, and the CPDS is fully restored. The overall process of the multi-stage service restoration is shown in Fig.4, which depicts the variation of the system function [27].

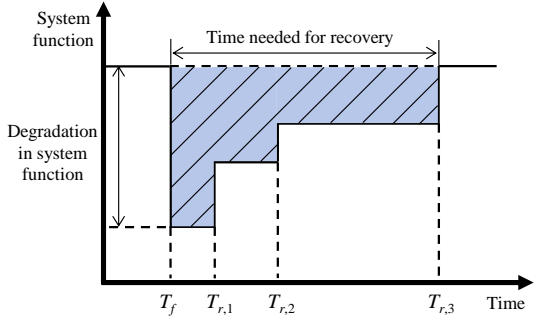

**Fig. 4.** The multi-stage service restoration process of the CPDS.

At $T_{r,1}$, the defenders leverage the RCSs to conduct the fault isolation and load transfer operation, subject to the following:

a) Topological constraints: the spanning tree method [28] ensures that the distribution network topology remains radial, expressed as:

$$\sum_{k\in\delta(j)} X_{kj,1} + \sum_{i\in\pi(j)} X_{ij,1} \leq 1 - g_j - dg_j\alpha_{j,1} + M\lambda_{j,1}, \forall j \in B \quad (10a)$$
$$\sum_{k\in\delta(j)} X_{kj,1} + \sum_{i\in\pi(j)} X_{ij,1} \geq 1 - g_j - dg_j\alpha_{j,1} - M\lambda_{j,1}, \forall j \in B \quad (10b)$$
$$X_{ij,1} + X_{ji,1} = c_{ij,1}, \forall ij \in O \quad (10c)$$

where $\pi(j)$ denotes the set of parent nodes of node $j$, $\delta(j)$ denotes the set of children nodes of node $j$, $B$ denotes the set of nodes, and $X_{ij,1}$ is a binary variable indicating the virtual flow direction of line $ij$, where $X_{ij,1} = 1$ if the virtual flow direction is $i \to j$ and $X_{ij,1} = 0$ otherwise. $g_j$ is a binary parameter indicating whether node $j$ is a substation node, $dg_j$ is a binary parameter indicating whether node $j$ is a distributed generation (DG) node, and $\alpha_{j,1}$ is a binary variable indicating whether the DG $j$ is the root bus in an island. The parameter $M$ represents a large number, and $\lambda_{j,1}$ is a binary variable indicating whether node $j$ belongs to the fault area. $c_{ij,1}$ is a binary variable indicating whether line $ij$ is closed. The subscripts 1 in the variables denotes the CPDS service restoration state at $T_{r,1}$.

The physical meaning of constraints (10a) and (10b) is that in the virtual network, which shares the same topology as the distribution network, each island has only one node injecting virtual flow, and each node receives at most one virtual flow. Constraint (10c) ensures that the virtual flow in a closed line is unidirectional and that no virtual flow exists in an open line.

b) Fault degradation constraints:

$$M\lambda_{j,1} \geq \sum_{i\in\pi(j)} s_{p,ij} + \sum_{k\in\delta(j)} c_{jk,1}s_{p,jk}, \forall j \in B \quad (11a)$$
$$\lambda_{i,1} - (1 - c_{ij,1}) \leq \lambda_{j,1} \leq \lambda_{i,1} + (1 - c_{ij,1}), \forall ij \in O \quad (11b)$$

where $s_{p,ij}$ is a binary parameter indicating whether a fault occurs on line $ij$. The physical meaning of constraint (11a) is that, under the assumption that the RCS is deployed at the head end of each line, when a power line fault occurs, the last node of the line must be in the fault area, and the status of the first node depends on the state of line $jk$. Constraint (11b) ensures that the nodes at both ends of a closed line must be in the same state, meaning they are either both in the non-fault area or both in the fault area.

c) Cyber-physical coupling constraints:

$$c_{ij,1} \geq s_{c,ij}, \forall ij \in O_1 \quad (12a)$$
$$c_{ij,1} \leq 1 - s_{c,ij}, \forall ij \in O_2 \quad (12b)$$

where $s_{c,ij}$ is a binary parameter indicating whether the RCS of line $ij$ is disconnected due to a FBS attack. $O_1$ denotes the set of power lines with segmented switches, and $O_2$ denotes the set of tie lines.

The physical meaning of constraint (12a) and (12b) is that if the RCS of line $ij$ is disconnected from the control center due to a FBS attack, the defenders cannot control the RCS, and the status of line $ij$ remains unchanged.

d) Operational constrains: The Lindistflow model [29] and the linearization method of second-order cone constraints [30] are used in the operation constrains of the service restoration model of the CPDS, and the details are given below:

$$\sum_{k\in\delta(j)} P_{jk,1} - \sum_{i\in\pi(j)} P_{ij,1} = P_{G,j,1} - (P_{L,j} - P_{S,j,1}), \forall j \in B \quad (13a)$$
$$\sum_{k\in\delta(j)} Q_{jk,1} - \sum_{i\in\pi(j)} Q_{ij,1} = Q_{G,j,1} - (Q_{L,j} - Q_{S,j,1}), \forall j \in B \quad (13b)$$
$$U_{i,1} - U_{j,1} \leq (1 - c_{ij,1}) \cdot M + 2(r_{ij}P_{ij,1} + x_{ij}Q_{ij,1}), \forall ij \in O \quad (13c)$$
$$U_{i,1} - U_{j,1} \geq -(1 - c_{ij,1}) \cdot M + 2(r_{ij}P_{ij,1} + x_{ij}Q_{ij,1}), \forall ij \in O \quad (13d)$$
$$U_{j,\min} \leq U_{j,1} \leq U_{j,\max}, \forall j \in O \quad (13e)$$
$$0 \leq P_{G,j,1} \leq P_{G,j,\max}, \forall j \in O \quad (13f)$$
$$0 \leq Q_{G,j,1} \leq Q_{G,j,\max}, \forall j \in O \quad (13g)$$
$$\lambda_{j,1}P_{L,j} \leq P_{S,j,1} \leq P_{L,j}, \forall j \in O \quad (13h)$$
$$\lambda_{j,1}Q_{L,j} \leq Q_{S,j,1} \leq Q_{L,j}, \forall j \in O \quad (13i)$$
$$-c_{ij,1}S_{ij,\max} \leq (\sqrt{2} - 1)P_{ij,1} + Q_{ij,1} \leq c_{ij,1}S_{ij,\max}, \forall ij \in O \quad (13j)$$
$$-c_{ij,1}S_{ij,\max} \leq (\sqrt{2} - 1)P_{ij,1} - Q_{ij,1} \leq c_{ij,1}S_{ij,\max}, \forall ij \in O \quad (13k)$$
$$-c_{ij,1}S_{ij,\max} \leq P_{ij,1} + (\sqrt{2} - 1)Q_{ij,1} \leq c_{ij,1}S_{ij,\max}, \forall ij \in O \quad (13l)$$
$$-c_{ij,1}S_{ij,\max} \leq P_{ij,1} - (\sqrt{2} - 1)Q_{ij,1} \leq c_{ij,1}S_{ij,\max}, \forall ij \in O \quad (13m)$$

where $P_{ij,1}$ and $Q_{ij,1}$ represent the active power and reactive power of line $ij$, $U_{j,1}$ denotes the square of the voltage magnitude at node $j$, $r_{ij}$ and $x_{ij}$ are the resistance and reactance of line $ij$, $P_{G,j,1}$ and $Q_{G,j,1}$ represent the active and reactive power outputs of the substation or DG at node $j$, $P_{L,j}$ and $Q_{L,j}$ denote the active and reactive loads at node $j$, $P_{S,j,1}$ and $Q_{S,j,1}$ denote the active and reactive load losses at node $j$, $P_{G,j,\max}$ and $P_{G,j,\max}$ denote the active and reactive limits of the substation or DG at node $j$, $S_{ij,\max}$ is the capacity limit of line $ij$, and $U_{j,\min}$ and $U_{j,\max}$ are the minimum and maximum values of the square of the voltage magnitude at node $j$.

At $T_{r,2}$, if some RCSs are disconnected from the control center due to FBS attack, the power grid defenders control the disconnected switches on-site to complete the fault isolation and load transfer operations. The constraints at this stage are similar to those at $T_{r,1}$, with the key difference that cyber-physical coupling constraints are no longer applicable, as all RCSs can be operated on-site.

$$\sum_{k\in\delta(j)} X_{kj,2} + \sum_{i\in\pi(j)} X_{ij,2} \leq 1 - g_j - dg_j\alpha_{j,2} + M\lambda_{j,2}, \forall j \in B \quad (14a)$$

$$\sum_{k\in\delta(j)} X_{kj,2} + \sum_{i\in\pi(j)} X_{ij,2} \geq 1 - g_j - dg_j\alpha_{j,2} - M\lambda_{j,2}, \forall j \in B \quad (14b)$$
$$X_{ij,2} + X_{ji,2} = c_{ij,2}, \forall ij \in O \quad (14c)$$
$$M\lambda_{j,2} \geq \sum_{i\in\pi(j)} s_{p,ij} + \sum_{k\in\delta(j)} c_{jk,2} s_{p,jk}, \forall j \in B \quad (14d)$$
$$\lambda_{i,2} - (1 - c_{ij,2}) \leq \lambda_{j,2} \leq \lambda_{i,2} + (1 - c_{ij,2}), \forall ij \in O \quad (14e)$$
$$\sum_{k\in\delta(j)} P_{jk,2} - \sum_{i\in\pi(j)} P_{ij,2} = P_{G,j,2} - (P_{L,j} - P_{S,j,2}), \forall j \in B \quad (14f)$$
$$\sum_{k\in\delta(j)} Q_{jk,2} - \sum_{i\in\pi(j)} Q_{ij,2} = Q_{G,j,2} - (Q_{L,j} - Q_{S,j,2}), \forall j \in B \quad (14g)$$
$$U_{i,2} - U_{j,2} \leq (1 - c_{ij,2}) \cdot M + 2(r_{ij}P_{ij,2} + x_{ij}Q_{ij,2}), \forall ij \in O \quad (14h)$$
$$U_{i,2} - U_{j,2} \geq -(1 - c_{ij,2}) \cdot M + 2(r_{ij}P_{ij,2} + x_{ij}Q_{ij,2}), \forall ij \in O \quad (14i)$$
$$U_{j,\min} \leq U_{j,2} \leq U_{j,\max}, \forall j \in O \quad (14j)$$
$$0 \leq P_{G,j,2} \leq P_{G,j,\max}, \forall j \in O \quad (14k)$$
$$0 \leq Q_{G,j,2} \leq Q_{G,j,\max}, \forall j \in O \quad (14l)$$
$$\lambda_{j,2}P_{L,j} \leq P_{S,j,2} \leq P_{L,j}, \forall j \in O \quad (14m)$$
$$\lambda_{j,2}Q_{L,j} \leq Q_{S,j,2} \leq Q_{L,j}, \forall j \in O \quad (14n)$$
$$-c_{ij,2}S_{ij,\max} \leq (\sqrt{2} - 1)P_{ij,2} + Q_{ij,2} \leq c_{ij,2}S_{ij,\max}, \forall ij \in O \quad (14o)$$
$$-c_{ij,2}S_{ij,\max} \leq (\sqrt{2} - 1)P_{ij,2} - Q_{ij,2} \leq c_{ij,2}S_{ij,\max}, \forall ij \in O \quad (14p)$$
$$-c_{ij,2}S_{ij,\max} \leq P_{ij,2} + (\sqrt{2} - 1)Q_{ij,2} \leq c_{ij,2}S_{ij,\max}, \forall ij \in O \quad (14q)$$
$$-c_{ij,2}S_{ij,\max} \leq P_{ij,2} - (\sqrt{2} - 1)Q_{ij,2} \leq c_{ij,2}S_{ij,\max}, \forall ij \in O \quad (14r)$$

where the subscript '2' in the variables denotes the CPDS recovery state at $T_{r,2}$.

*C. The tri-level optimization model*

The above three-stage DAD process can be modeled as a perfect information dynamic game, which can reach the Stackelberg equilibrium. The game is expressed by a triple $\Gamma = \langle I, S, u \rangle$, where $I = \{a, d\}$ is the player space, with $a$ denoting the attacker and $d$ denoting the defender. $S = \{S_a, S_d\}$ is the strategy space, where $S_a$ and $S_d$ represent the attacker's and defender's strategy spaces, respectively. $u_a(s_a, s_d)$ and $u_d(s_a, s_d)$ represent the payoffs for the attacker and defender, respectively, with $s_a \in S_a$ and $s_d \in S_d$. Since the results of the attacks are uncertain, the success probabilities of the attacks are modelled. For the attacker, there are two possible outcomes when a cyber or physical node is attacked, namely success or failure. If there are $N$ attacked cyber and physical nodes in total, the number of possible scenarios is $2^N$. To reflect the consequences of all possible scenarios, the payoffs for both the attacker and defender are designed as the expectation of the resilience of the CPDS under all possible scenarios, which can be expressed as:

$$u_d(s_a, s_d) = -u_a(s_a, s_d) = \max \sum_{h\in H} Prob_h \cdot R_h \quad (15a)$$

$$R_h = 1 - \begin{bmatrix}(T_{r,1} - T_f)\sum_{j\in B}\beta_j P_{S,j,0,h} \\ +(T_{r,2} - T_{r,1})\sum_{j\in B}\beta_j P_{S,j,1,h} \\ +(T_{r,3} - T_{r,2})\sum_{j\in B}\beta_j P_{S,j,2,h}\end{bmatrix} / [(T_{r,3} - T_f)\sum_{j\in B}\beta_j P_{L,j}] \quad (15b)$$

where $h$ represents the possible scenario, $H$ represents the set of all possible scenarios after the strategy of defender in Stage 1 and the strategy of attacker in Stage 2 are determined, $Prob_h$ is the probability of scenario $h$ under the determined strategies in Stage 1 and Stage 2, and $R_h$ represents the resilience of the CPDS in scenario $h$, as defined in (15b). $\beta_j$ is the weight of the load at node $j$, and $P_{S,j,0,h}$, $P_{S,j,1,h}$, $P_{S,j,2,h}$ represent the load losses at node $j$ at $T_f$, $T_{r,1}$, $T_{r,2}$, respectively. In a single feeder system, all loads on the feeder are lost at $T_f$, leading to:

$$P_{S,j,0,h} = P_{L,j}, \forall j \in B \quad (16)$$

The above three-stage perfect information dynamic game model can be transformed into a tri-level max-min-max optimization model, which can be expressed as follows:

$$\max_\alpha \min_\beta \max_\gamma \sum_{h\in H} Prob_h \cdot R_h \quad (17a)$$
$$\text{s.t.} (1) - (5), (9) - (14), (15b), (16)$$
$$Prob_h = \prod_{t\in\mathcal{E}_h} p_t(\alpha, \beta) \prod_{t\in\Xi_h}(1 - p_t(\alpha, \beta)) \quad (17b)$$

where $\alpha = \{D_{p,ij}, D_{c,w}\}$ represents the decision variables of the defender in Stage 1, $\beta = \{A_{p,ij}, X_{FBS}, Y_{FBS}\}$ represents the decision variables of the attacker in Stage 2, and $\gamma = \{c_{ij,1}, c_{ij,2}, P_{G,j,1}, P_{G,j,2}, Q_{G,j,1}, Q_{G,j,2}\}$ represents the decision variables of the defender in Stage 3. $p_t(\alpha, \beta)$ denotes the probability of a successful attack on the physical or cyber attack target $t$, $\mathcal{E}_h$ denotes the set of attack targets with successful outcomes in scenario $h$, and $\Xi_h$ denotes the set of attack targets with unsuccessful outcomes in scenario $h$.

III. SOLUTION METHODOLOGY

In the objective function of the tri-level optimization model, as shown in expression (17), the $Prob_h$ is only related to the decision variables $\alpha$ and $\beta$, and is independent of the decision variable $\gamma$. Therefore, the objective function (17a) can be expressed as:

$$\max_\alpha \min_\beta \sum_{h\in H} Prob_h \max_\gamma R_h \quad (18)$$

For each fault scenario, there exists a specific optimal fault recovery strategy $\gamma^*$, and the corresponding resilience $R_h^*$ can be determined. Thus, we fist solve the following mixed-integer linear programming (MILP) problem for each fault scenario to obtain $\gamma^*$ and $R_h^*$:

$$\max_\gamma R_h \quad (19)$$
$$\text{s.t.} (10) - (14), (15b), (16)$$

Next, the solved resilience $R_h^*$ is substituted into (18), transforming the tri-level optimization model into a bi-level optimization model consisting of a master problem (MP) and a subproblem (SP):

$$\max_\alpha \min_\beta \sum_{h\in H} Prob_h R_h^* \quad (20)$$
$$\text{s.t.} (1) - (5), (9), (17b)$$

The bi-level optimization model can be solved using C&CG method [31] combined with an enumeration method. The flowchart of the overall solution method is detailed in Fig.5, where $k$ is the number of iterations, $\varepsilon$ is the error condition for the convergence of the C&CG, and $\Lambda_k$ denotes the attack uncertainty scenario set of $k_{th}$ iteration in MP.

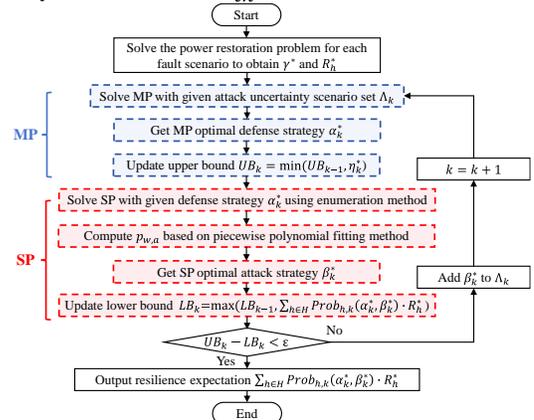

**Fig. 5.** The flowchart of the overall solution method.

## A. Subproblem (SP) formulation

$$\min_{\beta} \sum_{h \in H} Prob_{h,k} R_h^* \tag{21a}$$

$$\text{s.t.} (5), (9)$$

$$Prob_{h,k} = \prod_{t \in \mathcal{E}_h} p_t(\alpha_k^*, \beta) \prod_{t \in \Xi_h}(1 - p_t(\alpha_k^*, \beta)) \tag{21b}$$

$$p_t(\alpha_k^*, \beta) = p_{a,w}(1 - p_{d,w} D_{c,w,k}^*), t \in W \tag{21c}$$

$$p_t(\alpha_k^*, \beta) = A_{p,ij}\left(1 - \left(p_{ij,a} - D_{p,ij,k}^*(p_{ij,a} - p_{ij,b})\right)\right), t \in O \tag{21d}$$

where $\alpha_k^*$ represents the given defense strategy of $k_{th}$ iteration. In the SP formulation, the FBS attack probability $p_{a,w}$ is calculated based on (9), which is a transcendental integral constraint that cannot be directly calculated. To address this, an approximate calculation method based on piecewise polynomial fitting is proposed. Using this approximation, $p_{a,w}$ is computable but the SP formulation remains a highly nonlinear optimization problem, and since the attack strategy set is limited, the enumeration method is used for solution.

The transcendental integral formula (9) can be expressed in the following form:

$$\begin{aligned} p_{a,w} &= \int_{-\infty}^{+\infty} \phi(\tau) \prod_{k=1}^{N} \Phi\left(\frac{\sigma_{N+1}}{\sigma_k}\tau + \frac{\mu_{N+1} - \mu_N}{\sigma_k}\right) d\tau \\ &= \int_{-3}^{+3} z_{pdf}(\tau) \prod_{k=1}^{N} z_{cdf}\left(\frac{\sigma_{N+1}}{\sigma_k}\tau + \frac{\mu_{N+1} - \mu_N}{\sigma_k}\right) d\tau \end{aligned} \tag{22}$$

where $\phi(\tau) = \frac{1}{\sqrt{2\pi}} e^{-\frac{\tau^2}{2}}$ denotes the PDF of the standard normal distribution, $\Phi(\cdot)$ denotes the CDF of the standard normal distribution, $z_{pdf}(\cdot)$ denotes the piecewise polynomial approximation of $\phi(\cdot)$, and $z_{cdf}(\cdot)$ denotes the piecewise polynomial approximation of $\Phi(\cdot)$. Therefore, by using piecewise polynomials to separately fit the PDF and CDF of the standard normal distribution and substituting them into the integral in (22), an approximate explicit calculation method for $p_{a,w}$ can be derived. The piecewise seventh-order polynomial approximation functions for the PDF and CDF are expressed in (23) and (24), respectively.

$$z_{pdf}(x) = \begin{cases} -e_7 x^7 + e_6 x^6 - e_5 x^5 + e_4 x^4 - e_3 x^3 \\ \quad -e_2 x^2 - e_1 x + e_0, 0 < x < 3 \\ e_7 x^7 + e_6 x^6 + e_5 x^5 + e_4 x^4 + e_3 x^3 \\ \quad -e_2 x^2 + e_1 x + e_0, -3 < x < 0 \end{cases} \tag{23}$$

where $e_7 = 0.001577$, $e_6 = 0.0236$, $e_5 = 0.1311$, $e_4 = 0.3035$, $e_3 = 0.1221$, $e_2 = 0.4629$, $e_1 = 0.002786$, and $e_0 = 1$.

$$z_{cdf}(x) = \begin{cases} -h_7 x^7 + h_6 x^6 - h_5 x^5 + h_4 x^4 - h_3 x^3 \\ \quad + h_2 x^2 + h_1 x + h_{0+}, 0 < x < 6 \\ -h_7 x^7 - h_6 x^6 - h_5 x^5 - h_4 x^4 - h_3 x^3 \\ \quad - h_2 x^2 + h_1 x + h_{0-}, -6 < x < 0 \end{cases} \tag{24}$$

where $h_7 = 5.615 \times 10^{-5}$, $h_6 = 0.001385$, $h_5 = 0.01362$, $h_4 = 0.06604$, $h_3 = 0.1485$, $h_2 = 0.04542$, $h_1 = 0.3908$, $h_{0+} = 0.4999$, and $h_{0-} = 0.5001$.

***Proposition 1:***

The approximate explicit calculation method for $p_{a,w}$ presented in this paper has an upper bound for the error value relative to the true distribution given by $\frac{6(\zeta + N\kappa)}{\sqrt{2\pi}}$, where $\zeta$ and $\kappa$ are the piecewise polynomial fitting error bounds for $\phi(\cdot)$ and $\Phi(\cdot)$, respectively.

The proof of Proposition 1 is provided in the Appendix.

## B. Master problem (MP) formulation

The master problem formulation of the C&CG method can be expressed as:

$$\max_{\alpha} \eta \tag{25a}$$

$$\text{s.t.} (4)$$

$$\eta \geq \sum_{h \in H} Prob_{h,n} \cdot R_h^*, n \in \Lambda_k \tag{25b}$$

$$Prob_{h,n} = \prod_{t \in \mathcal{E}_h} p_t(\alpha, \beta_n^*) \prod_{t \in \Xi_h}(1 - p_t(\alpha, \beta_n^*)), n \in \Lambda_k \tag{25c}$$

$$p_t(\alpha, \beta_n^*) = p_{a,w,n}^*(1 - p_{d,w} D_{c,w}), t \in W, n \in \Lambda_k \tag{25d}$$

$$p_t(\alpha, \beta_n^*) = A_{p,ij,n}^*\left(1 - \left(p_{ij,a} - D_{p,ij}(p_{ij,a} - p_{ij,b})\right)\right) \tag{25e}$$

$$, t \in O, n \in \Lambda_k$$

where $\eta$ is the auxiliary variable for the objective function of MP, $n$ represents the attack scenario in the attack uncertainty scenario set $\Lambda_k$ of the $k_{th}$ iteration. $\beta_n^*$ represents the attack strategy for attack scenario $n$, and $p_{a,w,n}^*$ represents the possibility of the cyber target $t$ (the RCS $w$) being attacked by the FBS and receiving the strongest SIB signal from the FBS in attack scenario $n$. The variables in the form of $\cdot^*$, including $R_h^*$, $\beta_n^*$, $p_{a,w,n}^*$, and $A_{p,ij,n}^*$, represent the determined variables, which are treated as constants in the MP formulation.

By substituting (25d) and (25e) into (25c) and expanding (25c), it becomes evident that (25c) contains the products of multiple binary variables $D_{p,ij}$ and $D_{c,w}$. These products can be transformed into a MILP problem using the method of adding auxiliary variables, as described in (26). Suppose there is a product of $n$ binary variables $b_1, b_2, ..., b_n$, it can be replaced by the auxiliary variable $f$ that meets:

$$\begin{cases} f \leq b_1 \\ f \leq b_2 \\ \quad \dots \\ f \leq b_n \\ f \geq b_1 + b_2 + \cdots + b_n - (n-1) \end{cases} \tag{26}$$

## IV. CASE STUDY

In this study, a modified IEEE 33-node distribution network and a real-world 47-node distribution power system in China are elaborately investigated to validate the effectiveness of the proposed three-stage DAD dynamic game model and solution method. The topology and parameters of the two case study systems are detailed in [32]. The durations for stages 1, 2, and 3 of the recovery process of the CPDS are set to 1 minute, 30 minutes, and 30 minutes, respectively. The corresponding simulations are carried out in the MATLAB™ environment using the following hardware configuration: CPU: Intel i5.13600K, GPU: NVIDIA RTX 2080Ti, RAM: 64G DDR43600MHz.

### A. Modified IEEE 33-node distribution network

The modified IEEE 33-node distribution network contains 5 DGs connected to nodes 6, 18, 21, 24, and 33. The RCSs are located at the head end of each line, and there are 7 legitimate BSs in the region, whose positions are illustrated in Fig.10.

1) Attack and defense strategy analysis under different resource scenarios

In this section, a detailed analysis of attack and defense strategies across various scenarios with different levels of attack and defense resources is given. Fig.6 illustrates 27 distinct scenarios, where cyber defense resources range from 2 to 4, physical defense resources range from 2 to 4, and physical attack resources range from 1 to 3.

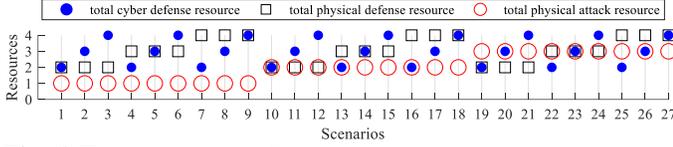

**Fig. 6.** The attack and defense resource scenarios.

The attack and defense strategies under the 27 scenarios are shown in Fig.7, where a blue solid rectangle denotes that physical defense is allocated to a power line, the red solid rectangle denotes that a physical attack is launched to a power line, a blue hollow rectangle denotes that cyber defense is allocated to the RCS of a power line, and a red hollow rectangle denotes that the RCS of a power line may be affected by a FBS attack. Given the large number of RCSs influenced by the FBS attack, only the three RCSs with the highest probability of being attacked are displayed in Fig.7.

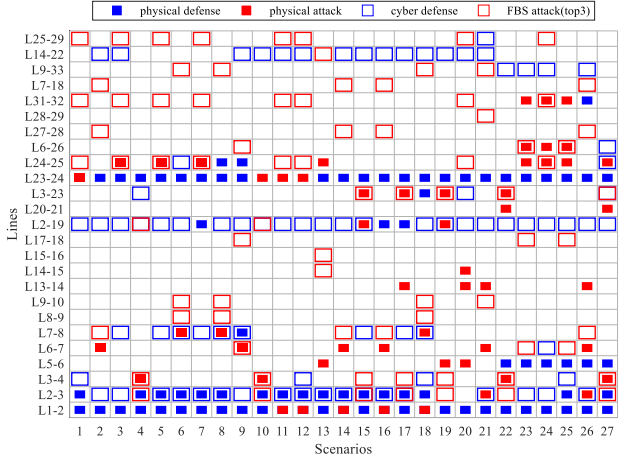

**Fig. 7.** The attack and defense strategies under different scenarios (IEEE 33-node distribution network).

From Fig.7, it is evident that physical defense resources are preferentially deployed on the power lines connected to or near substations and large DGs, such as L1-2, L2-3, and L23-24. This pattern also applies to physical attacks, where rational attackers, under the assumption of a complete information dynamic game, target unprotected power lines. On the cyber side, the defense resources are allocated to the RCSs of the power lines connected to high-degree nodes, such as L2-19, L2-3, and L14-22, to isolate faults and prevent their spread to the fault-free branches.

2) The resilience analysis under different cyber-physical attack and defense resources

The resilience of the IEEE 33-node distribution network under different cyber-physical defense resources is shown in Fig.8, with the physical attack resource ($N_{a,p} = 3$) and the reference signal strength ($S(d_0) = 100$). In Fig.8, the x-axis represents cyber defense resource $N_{d,c}$, the y-axis represents physical defense resource $N_{d,p}$, and the z-axis represents the resilience index of the distribution network. The results demonstrate that the resilience increases with the enhancement of both physical and cyber defense resources. When $N_{d,c} = 4$ and $N_{d,p} = 4$, the resilience under coordinated attacks improves from 46.73% to 72.01% compared to the case without any defense.

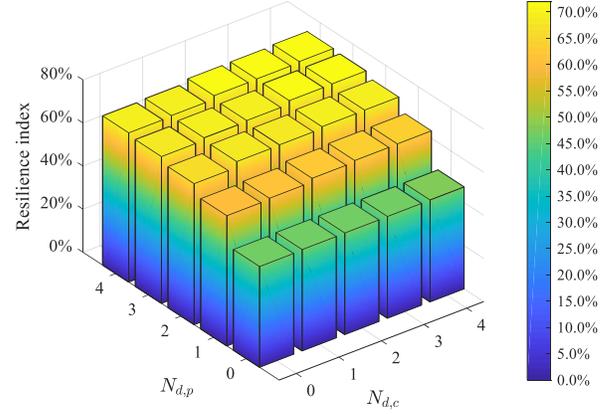

**Fig. 8.** The resilience under different cyber-physical defense resources (IEEE 33-node distribution network).

The resilience loss under different physical attack resources and FBS power levels is shown in Fig.9, with $N_{d,p} = 2$, $N_{d,c} = 2$, physical attack resource ranging from 1 to 3, and reference signal strength $S(d_0)$ ranging from 90 to 110. Scenarios without cyber attacks are also included for comparison. It is observed that increased investment in physical attack resources leads to higher resilience loss. For example, with $S(d_0) = 100$, the resilience losses for $N_{a,p} = 1, 2, 3$ are 11.74%, 20.79%, and 31.74% respectively. The presence of FBS further increases the resilience loss. When $N_{a,p} = 1, 2, 3$, the resilience losses under the $S(d_0) = 110$ are 1.52, 1.47, and 1.22 times higher, respectively, than in scenarios without FBS. As FBS power increases, the probability of hijacking nearby RCSs rises, leading to a greater resilience loss. However, this also increases the risk of detection for cyber attackers.

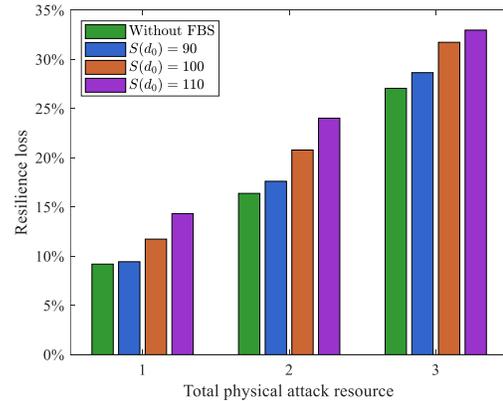

**Fig. 9.** The resilience loss under different physical attack resource and FBS power levels (IEEE 33-node distribution network).

3) The optimal FBS locations under different reference signal strengths

The optimal FBS locations under different reference signal strengths ($S(d_0) = 100, 104, 108, 112$) are shown in Fig.10.

In Fig.10, a black hollow circle denotes the normal unprotected RCS, a green circle denotes the protected RCS, a red triangle denotes the FBS, and a blue triangle denotes the BS. When the reference signal strength of the FBS is relatively low (100 and 104), the cyber attackers tend to position the FBS as close as possible to the critical RCSs to ensure a higher probability of successful attacks. Conversely, when the reference signal strength of the FBS is relatively high (108 and 112), the probability of attack success remains high even when the FBS is located at a greater distance from the RCSs. In this case, the attackers tend to position the FBS at a central geographical location, equidistant from several switches, to enable the FBS to target more RCSs.

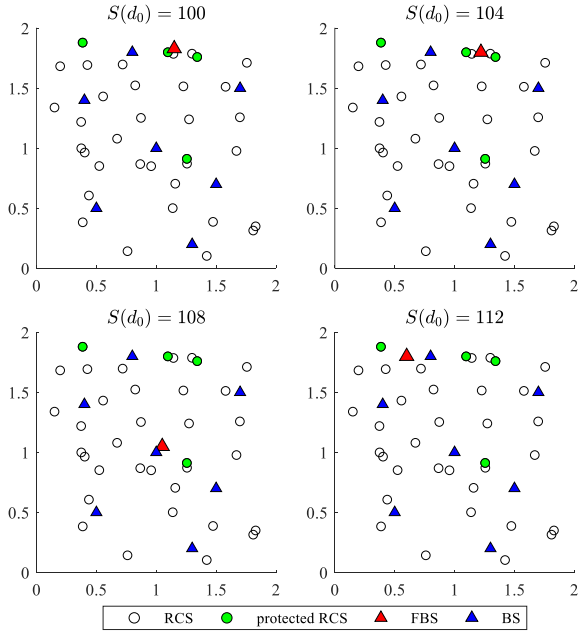

**Fig. 10.** The optimal FBS locations under different reference signal strengths (IEEE 33-node distribution network).

*B. A real-world 47-node distribution power system in China*

To further validate the effectiveness of the proposed model and method, simulation studies are performed on a real-world 47-node distribution power system in China. The topology of the system is shown in Fig.11.

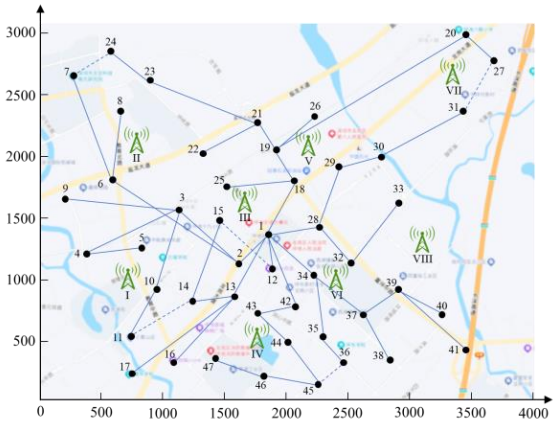

**Fig. 11.** The topology of the real-world 47-node distribution power system in China.

Similarly, the 27 scenarios depicted in Fig.6 are studied for the real-world 47-node distribution power system, and the corresponding attack and defense strategies are given in Fig.12. From Fig.12, conclusions analogous to those of the IEEE 33-node distribution network can be drawn: (i) physical defense resources are preferentially deployed on power lines connected to or near substations and large DGs, such as L1-34, L1-18, L1-42, and L34-37; and (ii) cyber defense resources are allocated to the RCSs of lines connected to high-degree nodes, such as L1-28, L1-13, L1-2, and L1-42.

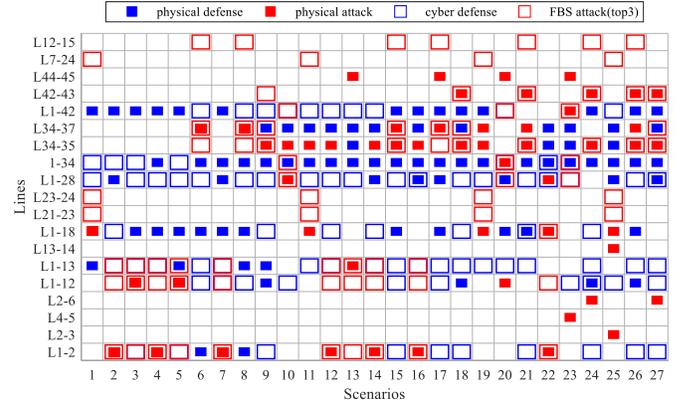

**Fig. 12.** The attack and defense strategies under different scenarios (real-world 47-node distribution power system).

The resilience level of the 47-node distribution power system under different cyber-physical defense resource is shown in Fig.13, with physical attack resource ($N_{a,p} = 3$) and reference signal strength ($S(d_0) = 100$). The results demonstrate that increasing both physical and cyber defense resources significantly enhances the resilience of system. Specifically, the resilience improves from 42.19% (without defense) to 72.16% when $N_{d,p} = 4$ and $N_{d,c} = 4$.

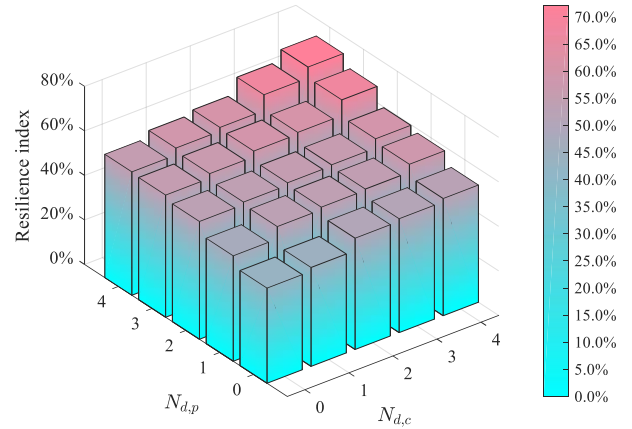

**Fig. 13.** The resilience under different cyber-physical defense resources (real-world 47-node distribution power system).

## V. CONCLUSION

In this paper, a novel resilience assessment framework for CPDSs is proposed, addressing coordinated cyber-physical attacks involving FBS and physical attacks within a DAD dynamic game framework. The principles and implementation process of FBS attack in CPDSs are analyzed elaborately, and a FBS attack model is established based on the lognormal shadowing model. The multi-stage service restoration process, incorporating cyber-physical coupling, is integrated into the service restoration model to account for the impact of attacked RCSs on the physical-side recovery process. The constructed DAD dynamic game model is equivalently recast as a tri-level max-min-max optimization model, which is solved using C&CG technique combined with enumeration method. Simulations performed on the modified IEEE 33-node CPDS and a real-world 47-node CPDS in China demonstrate that analysis of attack and defense strategies under the dynamic game framework highlights the critical importance of power lines connected to or near substations and DGs, as well as RCSs connected to high-degree nodes. In addition, the optimal positioning of the FBS varies with signal power. In low signal power scenarios, attackers position the FBS as close as possible to critical RCSs, whereas in high signal power scenarios, they prefer a central geographical location equidistant from multiple switches.

### APPENDIX: THE PROOF OF PROPOSITION 1

*Remark1:*

Based on the 3-sigma rule of the normal distribution, equation (9) can be approximated as:

$$p_{a,w} = \frac{1}{\sqrt{2\pi}} \int_{-3}^{+3} e^{-\frac{\tau^2}{2}} \prod_{k=1}^{N} \Phi\left(\frac{\sigma_{N+1}}{\sigma_k}\tau + \frac{\mu_{N+1}-\mu_k}{\sigma_k}\right) d\tau$$
$$= \frac{1}{\sqrt{2\pi}} \int_{-3}^{+3} e^{-\frac{\tau^2}{2}} \prod_{k=1}^{N} \Phi(\tau + a_k) d\tau \quad (27)$$

where $a_k = \frac{\mu_{N+1}-\mu_k}{\sigma_k}$, and within a certain nearby range, it is assumed that the variance of the normal distribution introduced by the shadowing effect is consistent, i.e., $\frac{\sigma_{N+1}}{\sigma_k} = 1$.

*Remark2:*

(a) Define $V$ as a linear space, then:
$$\forall \boldsymbol{\alpha}, \boldsymbol{\beta} \in V, \|\boldsymbol{\alpha}+\boldsymbol{\beta}\|_1 \leq \|\boldsymbol{\alpha}\|_1 + \|\boldsymbol{\beta}\|_1 \quad (28)$$

(b) Triangle inequality for integrals: let $f(t)$ be a continuous and integrable function on the closed interval $[a,b]$, and then:
$$\left|\int_a^b f(t)\,dt\right| \leq \int_a^b |f(t)|\,dt \quad (29)$$

(c) $\forall k \in N$, $f_k(t)$ is a continuous and integrable function on the closed interval $[a,b]$. The function $z_k(t)$ is the polynomial approximation of $f_k(t)$, and the error function is defined as $\varepsilon_k(t) = f_k(t) - z_k(t)$. Let $M_k = \max_{t \in [a,b]} |f_k(t)|$. Define $F(t) = \prod_{k=1}^{N} f_k(t)$ and $Z(t) = \prod_{k=1}^{N} z_k(t)$ The following properties can be obtained:

$$F(t) - Z(t) = \prod_{k=1}^{N} f_k(t) - \prod_{k=1}^{N} z_k(t)$$
$$= \prod_{k=1}^{N}(z_k(t) + \varepsilon_k(t)) - \prod_{k=1}^{N} z_k(t) \quad (30)$$

By ignoring the higher-order terms of $\varepsilon_k(t)$, we obtain:
$$F(t) - Z(t) \approx \sum_{k=1}^{N} \varepsilon_k(t) \prod_{i \neq k} z_i(t) \quad (31)$$

From Remark 2 (a), it follows that:
$$|F(t) - Z(t)| \leq \sum_{k=1}^{N} |\varepsilon_k(t)| |\prod_{i \neq k} z_i(t)|$$
$$\leq \sum_{k=1}^{N} |\varepsilon_k(t)| \prod_{i \neq k} M_i \quad (32)$$

*Proof:*

According to Remark 1, during the polynomial fitting process for the PDF function of the standard normal distribution, only the interval of the random variable from -3 to 3 needs to be considered. However, for the CDF function, due to the existence of $a_k \in (-3,3)$, the interval of the random variable from -6 to 6 should be accounted for during its polynomial fitting process. The fitting error for the PDF and CDF of the standard normal distribution are shown in Fig.14 and Fig.15, respectively. Based on Monte Carlo sampling, the error bounds resulting from the piecewise polynomial fitting are $\zeta = 2.678 \times 10^{-4}$ for the PDF and $\kappa = 5.69 \times 10^{-4}$ for the CDF.

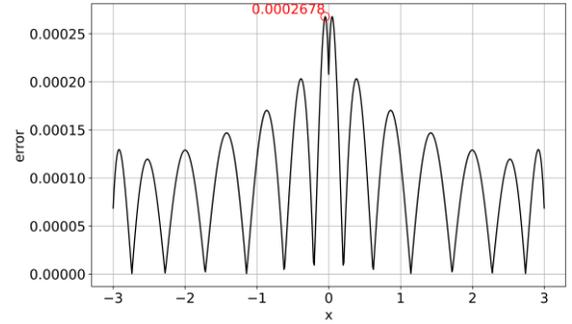

**Fig. 14.** The fitting error of PDF of the standard normal distribution

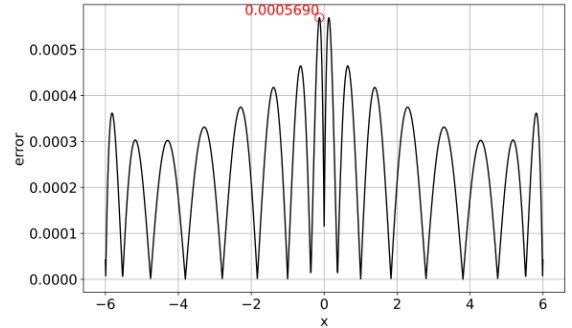

**Fig. 15.** The fitting error of CDF of the standard normal distribution.

Given a Gaussian distribution $\mathcal{N}_m(t|\mu_m, \sigma_m^2)$ with mean $\mu_m$ and variance $\sigma_m^2$, the error between its CDF and the piecewise polynomial fitting function can be derived as follows:

$$\left|\int_{-\infty}^{x} \mathcal{N}_m(t|\mu_m, \sigma_m^2)\,dt - z_{cdf}\left(\frac{x-\mu_m}{\sigma_m}\right)\right|$$
$$= \left|\frac{1}{\sqrt{2\pi}} \int_{-\infty}^{\frac{x-\mu_m}{\sigma_m}} e^{-\frac{(t-\mu_m)^2}{2\sigma_m^2}} d\left(\frac{t}{\sigma_m}\right) - z_{cdf}\left(\frac{x-\mu_m}{\sigma_m}\right)\right| \quad (33)$$

Two new variables as defined as follows:
$$\tilde{t} = \frac{t-\mu_m}{\sigma_m}, \tilde{x} = \frac{x-\mu_m}{\sigma_m} \quad (34)$$

Substituting (34) into (33) yields:
$$= \left|\frac{1}{\sqrt{2\pi}} \int_{-\infty}^{\tilde{x}} e^{-\frac{\tilde{t}}{2}} d\tilde{t} - z_{cdf}(\tilde{x})\right|$$
$$= |\Phi(\tilde{x}) - z_{cdf}(\tilde{x})| \leq \kappa \quad (35)$$

Equation (35) represents that for the Gaussian distribution $\mathcal{N}_m(t|\mu_m, \sigma_m^2)$, the error bound with respect to the piecewise polynomial fitting function remains less than $\kappa$.

Based on the above, returning to focus on (9), let $\tilde{\tau} = \tau + a_k$. The error upper bound of the proposed piecewise polynomial fitting approximation method for (9) can be derived as follows:

$$|p_{a,w} - z_{fit}|$$
$$= \frac{1}{\sqrt{2\pi}}\left|\int_{-3}^{+3} e^{-\frac{\tau^2}{2}} \prod_{k=1}^{N} \Phi(\tilde{\tau}) d\tau - \int_{-3}^{+3} z_{pdf}(\tau) \prod_{k=1}^{N} z_{cdf}(\tilde{\tau}) d\tau\right| \quad (36)$$

From the integral form of the triangle inequality in Remark 2 (b), it follows that:

$$= \frac{1}{\sqrt{2\pi}}\left|\int_{-3}^{+3} e^{-\frac{\tau^2}{2}} \prod_{k=1}^{N} \Phi(\tilde{\tau}) d\tau - \int_{-3}^{+3} z_{pdf}(\tau) \prod_{k=1}^{N} z_{cdf}(\tilde{\tau}) d\tau\right|$$
$$\leq \frac{1}{\sqrt{2\pi}}\int_{-3}^{+3}\left|e^{-\frac{\tau^2}{2}} \prod_{k=1}^{N} \Phi(\tilde{\tau}) - z_{pdf}(\tau) \prod_{k=1}^{N} z_{cdf}(\tilde{\tau})\right| d\tau \quad (37)$$

In addition, from Remark 2 (c), the above expression can be further derived as:

$$\leq \frac{1}{\sqrt{2\pi}}\int_{-3}^{+3}|\sum_{k=1}^{N} \varepsilon_k(\tau)| \prod_{i \neq k} M_i | d\tau \quad (38)$$

where $\varepsilon_k(\tau)$ represents the error function between the piecewise seventh-order polynomial function and the original function, and $M_i$ is the maximum absolute value of the original function (the CDF and PDF of the standard normal distribution), i.e., $\forall i \in N, M_i = 1$. Based on Monte Carlo sampling, the upper bounds of the error function for the CDF and PDF of the standard normal distribution are:

$$|\varepsilon_{pdf}(\tau)| \leq \zeta, |\varepsilon_{cdf}(\tau)| \leq \kappa \quad (39)$$

Therefore, it can be concluded that:

$$|p_{a,w} - z_{fit}| \leq \frac{1}{\sqrt{2\pi}}\int_{-3}^{+3}|\sum_{k=1}^{N} \varepsilon_k(\tau)| \prod_{i \neq k} M_i | d\tau$$
$$\leq \frac{1}{\sqrt{2\pi}}\int_{-3}^{+3}|\zeta + N\kappa| d\tau = \frac{6(\zeta + N\kappa)}{\sqrt{2\pi}} \quad (40)$$